\documentclass[graybox, nosecnum]{svmult}

\usepackage{mathptmx}       
\usepackage{helvet}         
\usepackage{courier}        
\usepackage{type1cm}        
\usepackage{makeidx}         
\usepackage{graphicx}        
\usepackage{multicol}        
\usepackage[bottom]{footmisc}
\usepackage{hyperref}        
\usepackage{soul}            
\hypersetup{colorlinks=true,urlcolor=blue}
\usepackage{url}
\usepackage{multirow}
\usepackage[dvipsnames]{xcolor}
\usepackage[square,numbers]{natbib}
\newcommand{\hbindex}[1]{\hl{#1}\index{#1}}  
\makeindex             

\begin{document}
\title*{Pulsar Timing Array Experiments}
\author{J.~P.~W.~Verbiest\thanks{corresponding author},
  S.~Os{\l}owski and S.~Burke-Spolaor} 
\institute{J.~P.~W.~Verbiest \at Fakult\"at f\"ur Physik,
  Universit\"at Bielefeld, Postfach 100131, 33501 Bielefeld, Germany
  and\\ Max-Planck-Institut f\"ur Radioastronomie, Auf dem H\"ugel 69,
  53121 Bonn, Germany, \email{verbiest@physik.uni-bielefeld.de}
\and S.~Os{\l}owski \at Centre for Astrophysics and Supercomputing,
Swinburne University of Technology, PO Box 218, Hawthorn, VIC 3122,
Australia, \email{stefan.oslowski@gmail.com}
\and S.~Burke-Spolaor \at Department of Physics and Astronomy, West
Virginia University, P.O.\ Box 6315, Morgantown, WV 26506, USA and\\
Center for Gravitational
Waves and Cosmology, West Virginia University, Chestnut Ridge Research
Building, Morgantown, WV 26505, USA and\\ Canadian Institute for
Advanced Research, CIFAR Azrieli Global Scholar, MaRS Centre West
Tower, 661 University Ave.\ Suite 505, Toronto ON M5G 1M1, Canada,
\email{sarahbspolaor@gmail.com}} 
%
%
\maketitle
\abstract{ Pulsar timing is a technique that uses the highly stable
  spin periods of neutron stars to investigate a wide range of topics
  in physics and astrophysics. 
  Pulsar timing arrays (PTAs) use sets of extremely well-timed pulsars
  as a Galaxy-scale detector with arms extending between Earth and
  each pulsar in the array. These challenging experiments look for
  correlated deviations in the pulsars' timing that are caused by
  low-frequency gravitational waves (GWs) traversing our Galaxy.
  PTAs are particularly sensitive to GWs at nanohertz
  frequencies, which makes them complementary to other space-
  and ground-based detectors. In this chapter, we will describe the
  methodology behind pulsar timing; provide an overview of the
  potential uses of PTAs; and summarise where current PTA-based
  detection efforts stand. Most predictions expect PTAs to
  successfully detect a cosmological background of GWs emitted by
  supermassive black-hole binaries and also potentially detect continuous-wave emission from binary supermassive black holes, within the next several
  years.}

\section{Keywords} 
Pulsars; Pulsar Timing; Timing Array; Black Holes

\section{Introduction}

\begin{figure}[]
  \includegraphics[scale=0.5,angle=-90]{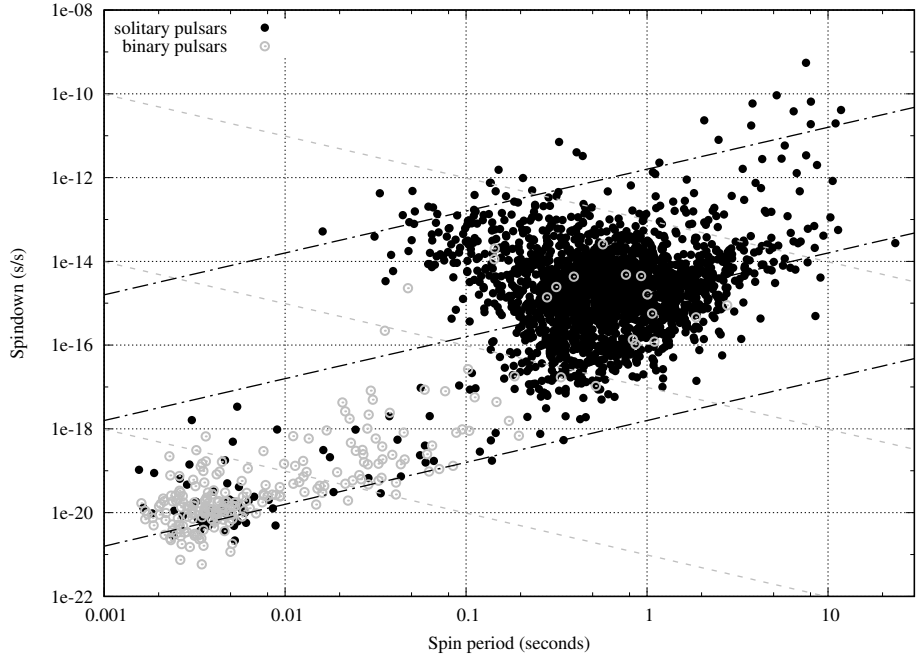}
  \caption{$P-\dot{P}$ diagram for all pulsars known. Shown are the
    spin period and spin-period derivative for all pulsars included in
    the ATNF pulsar catalogue version 1.61 \citep{mhth05}. Solitary
    pulsars are shown as full dots and pulsars in binary systems are
    shown as open circles. The grey dotted lines slanting downwards
    from the left represent surface magnetic-field strengths of
    $10^{13}$\,G, $10^{11}$\,G and $10^9$\,G from top to bottom; and
    the black dot-dashed lines slanting upwards to the right represent
    characteristic ages of $10^4$\,yr, $10^8$\,yr and $10^{10}$\,yr
    respectively, also from top to bottom.}
  \label{fig:ppdot}     
\end{figure}

\setcounter{footnote}{0}
\index{neutron star} Neutron stars are the collapsed cores of massive
stars that have undergone a supernova explosion after the end of
nuclear burning and are supported from further collapse by neutron
degeneracy pressure \citep{bz34c,gol68,pac68}. Since neutron stars are
far more compact than their progenitor stars, they tend to exhibit
very short rotational periods and extremely strong magnetic fields,
as shown in Figure~\ref{fig:ppdot}.
Generally the magnetic axis is not aligned with the spin axis, so
magnetic dipole radiation that is created in the neutron star's
atmosphere is swept around in space, somewhat like the beam of a
lighthouse (this is the so-called ``\hbindex{lighthouse
  model}''). Depending on the orientation of the beam and its width,
Earth may fall within that radiation beam once per
rotation -- which then causes the neutron star to be detected as a
source of pulsed radiation, otherwise referred to as a ``\hbindex{pulsar}''.

\subsection{Radio Emission from Pulsars}

\begin{figure}[]
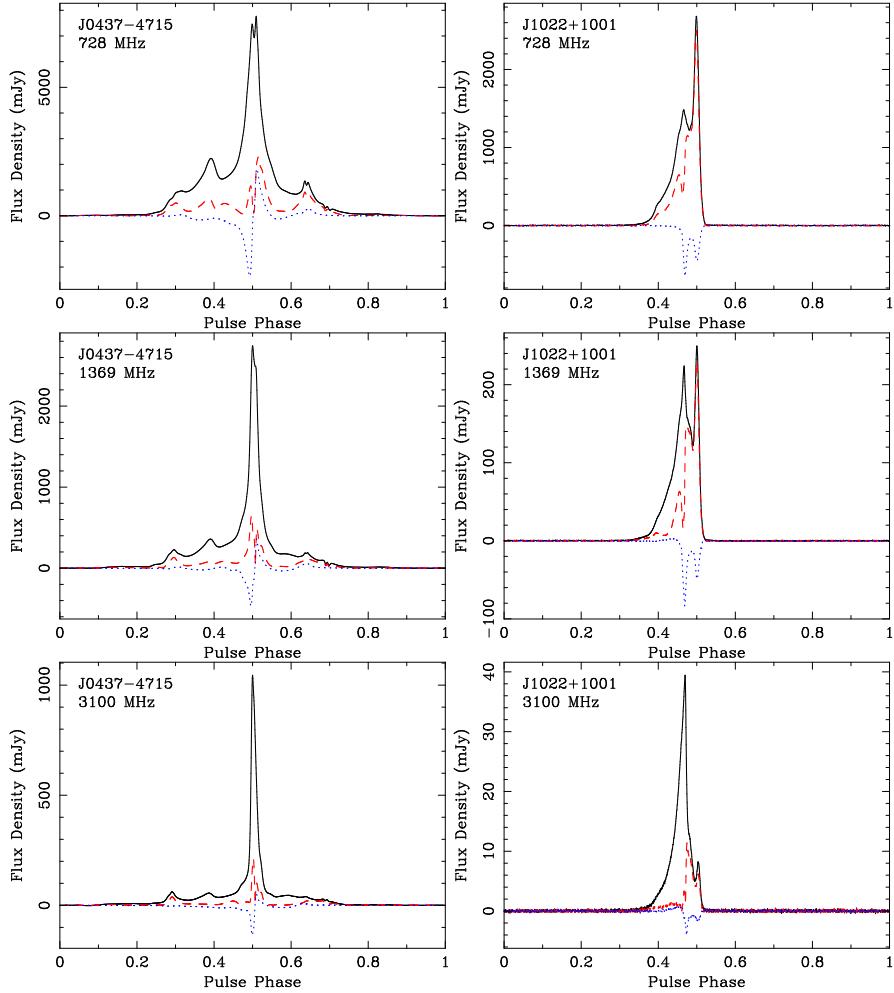

  \includegraphics[height=0.5\textwidth,angle=-90]{J0437-4715_50cm.ps}
  \includegraphics[height=0.5\textwidth,angle=-90]{J1022+1001_50cm.ps}
  \includegraphics[height=0.5\textwidth,angle=-90]{J0437-4715_20cm.ps}
  \includegraphics[height=0.5\textwidth,angle=-90]{J1022+1001_20cm.ps}
  \includegraphics[height=0.5\textwidth,angle=-90]{J0437-4715_10cm.ps}
  \includegraphics[height=0.5\textwidth,angle=-90]{J1022+1001_10cm.ps}
  \caption{Pulse profile shapes vary across pulsars and observing
    frequencies. Shown are pulse profiles for two traditional PTA
    pulsars: PSR~J0437$-$4715 (left column), and PSR~J1022+1001 (right
    column); and for each pulsar
    this profile is shown at three different observing frequencies: at
    $\sim726\,$MHz (top row), $\sim1369\,$MHz (middle row) and
    $\sim3100\,$MHz (bottom row). Total intensity profiles are shown
    in full black lines, linear polarisation in dashed red lines and
    circular polarisation in blue dotted lines. The tendency of
    getting sharper profile shapes at higher frequencies causes the
    timing precision to increase at those frequencies, although
    generally the noise level also increases due to the steep spectrum
    of pulsars \citep{blv13,jvk+18}. Consequently, most pulsar timing
    to date has been carried out at intermediate frequencies around
    1.4\,GHz. Finally, while this figure only shows profiles for MSPs,
    the evolution with frequency has been shown to be far more extreme
    in the case of slow pulsars \citep{kxl+98}. (These plots were made
    based on the public data published by \citet{dhm+15} and contain
    hundreds of hours of observing time at the higher frequencies,
    causing the radiometer noise to be barely visible.)}
  \label{fig:profs}
\end{figure}

Following the lighthouse model described above, it is natural to
expect that pulsars appear to the observer as so-called ``pulse
trains'': pulses of emission separated by a fixed period that equals
the spin.  These appear with a shape defined by the plasma properties
in the pulsar's magnetosphere, which can differ greatly from one
pulsar to the next (see Figure~\ref{fig:profs}). The emission
mechanism of pulsars is understood in broad terms \citep[see
  e.g.,][]{ml99b}. A few intriguing observational facts have been
identified over the half century since the first pulsars were
discovered. Most importantly, it has been shown that for most pulsars,
the exact shape of individual pulses changes randomly from one period
to the next. In contrast, however, the \emph{average} shape of the
pulsed emission is typically stable on timescales from minutes up to
decades \citep{hmt75}. This implies that for any given pulsar, the
pulsed emission can be averaged after accounting for the pulsar's
rotational period. The average pulse shape that can be obtained in
this way is unique for the pulsar and can be thought of as a
fingerprint. At radio wavelengths this average and reproducible pulse
shape is typically called the \hbindex{pulse profile} of the
pulsar, whereas the term \hbindex{light curve} is more common at higher frequencies (gamma and X-rays). The shape of the
profile is defined by the emitting geometry in the pulsar
magnetosphere. Since it is expected that the emission height is
different for photons with different frequencies \citep{cor78}, it
stands to reason that the shape of the pulse profile also typically
differs with observing frequency (again, see Figure~\ref{fig:profs}).

\index{pulsars!radio phenomenology}Not all pulsars have a stable pulse
profile and not all pulsars emit radiation all the time. Indeed, a
veritable zoo of pulsar emission phenomena has been discovered,
studied and described throughout the
years. \index{pulsars!nulling}There are so-called ``nulling'' pulsars 
\citep{bac70}, which often turn off, only to reappear at some point
after. Some pulsars null for minutes on end, others for hours -- some
even turn off for months or years (at which point they are also called
``intermittent'' pulsars), suggesting an almost continuous
distribution all the way up to \index{pulsars!rotating radio
  transients}so-called ``RRATs'' (Rotating Radio Transients), which
only sporadically emit one or several pulses of radiation 
\citep{mll+06,bb10}. Another category of pulsar emission is displayed
by the ``moding'' \index{pulsars!moding}pulsars \citep{bac70a}. These
do not have one characteristic fingerprint, but two or three -- and they
arbitrarily change between them: while one day their pulse profile may
look one way, the next day it may look different, only to go back to
its original state on day three. Moding can also have a wide range of
possible time scales, from single pulses all the way up to months or
years between mode changes. Finally, there are drifting pulsars 
\citep{dc68}. These ``drifters'' also have a well-defined pulse period
that is readily and repetitively measurable on timescales of minutes
to hours, but on timescales of seconds that pulse period seems to be
overestimated, as the pulse appears to come a bit too late after each
rotation, only to ``reset'' after a few dozen rotations, leading to
the more stable long-term periodicity.\index{pulsars!drifting
  sub-pulses}

Luckily for pulsar astronomers, the nulling, moding and drifting
pulsars have turned out to be the exception rather than the rule and
the majority of pulsars manifest themselves as predictably repetitive
pulses of emission that have arbitrary pulse shapes on timescales of
seconds, but stable and well-defined pulse shapes on timescales from
minutes upwards.

\subsection{Pulsar Lifecycle and Spin Properties}

Whereas the first few pulsars that were discovered appeared to be a
fairly homogeneous group of objects, extensive surveys have
continuously expanded the parameter space in which pulsars have been
discovered; and consequently a wide variety of pulsar types is now
known.

After the formative supernova explosion, pulsars start their new life
as so-called \hl{``young'' pulsars}\index{pulsars!young pulsar} with
spin periods of a few tens of milliseconds and a magnetic field
strength of order $10^{12}$\,G at their surface 
\citep{st83,bcdm03}. The emission of magnetic dipole radiation does
make them lose angular momentum and consequently their rotation slows
down gradually, typically by about $10^{-13}$\, seconds per
second. The youngest and most well-known example of this class of
pulsars is the Crab pulsar which was formed in a supernova explosion
in 1054 CE \citep{lun21,sr68}, which has been recorded by several
civilisations across the World.\index{pulsars!Crab pulsar}

After the first few thousand years, the spin period of young pulsars
has slowed down sufficiently to have an appreciable impact on the
spindown itself, which slows down their evolution. At this point they
turn into the first discovered type of pulsar: the so-called
\hl{``slow'' pulsars}\index{pulsars!slow pulsar} or \hl{``canonical''
  pulsars}\index{pulsars!canonical pulsar}. These are pulsars with
rotation periods between about a tenth of a second and roughly ten
seconds. Their magnetic fields are thought to have strengths of
roughly $10^{10}$\,G to $10^{13}$\,G and they are expected to have
formed thousands to hundreds of millions of years ago (see
Figure~\ref{fig:ppdot}). Their spin period gets longer by about
$10^{-14}$\, s every second, as they lose angular momentum slowly but
steadily. This loss in angular momentum causes them to eventually
rotate too slowly to produce detectable amounts of radio waves and so
after about a billion years they ``turn off'' and become undetectable 
\citep{rr94}.

Most of the slow pulsars are solitary objects that were either born as
solitary stars, or broke free from their companion stars during their
supernova explosion. A small sub-set, however, do have companion stars
that are almost without exception main-sequence
stars. \index{pulsars!binary pulsars}When these main-sequence stars
evolve and turn into red giants, it is not uncommon that their outer
atmospheric layers stray into the gravitational well of the pulsar and
cause it to accrete matter; and along with the matter, angular
momentum. These pulsars are then spun-up while their magnetic field
decays. The result is a \index{pulsars!millisecond
  pulsar}\hl{``millisecond'' pulsar (MSP)}\index{millisecond
  pulsar}\index{msp@MSP|see {millisecond pulsar}}, with spin periods
between 1\,ms and about 30\,ms. Due to the accretion process, MSPs
have relatively weak magnetic fields ($10^9$\,G or less) and
consequently their spin-down rates are also far lower than for slow
pulsars (typically of the order of $10^{-20}$\,s/s). As a result the
rotational and emission properties of MSPs are not thought to
appreciably evolve over their lifetimes. For an overview of formation
and evolution of MSPs see \citep{acrs82,bv91}.

Given the extremely small spin periods, stable pulse profiles for MSPs
can already be obtained in a matter of minutes or even
seconds. Furthermore, to date, only very few MSPs \citep{mvmp18} have
been demonstrated to show any of the anomalous emission properties
(nulling, moding, drifting) mentioned in the previous section that
some of the slow pulsars display. Finally, due to the much larger
angular momentum, MSPs have turned out to be far more stable clocks
than slow pulsars. These are the reasons why MSPs have become known as
``nature's gift to physics'': the perfect Einstein clock that can be
used to test a wide range of relativistic predictions.

\section{Pulsar Timing and Pulsar Timing Arrays}
\label{sec:2}

\hl{Pulsar timing}\index{pulsar timing} is a method that exploits the
highly regular spin period of pulsars and their predictable pulse
shapes, to study a wide range of questions in physics and
astrophysics. In essence, when doing pulsar timing, one monitors the
times at which subsequent pulses from a pulsar arrive at an
observatory. These observed \hl{pulse-arrival
  times}\index{pulse-arrival time} or \hl{ToAs}\index{toa@ToA|see
  {pulse-arrival time}} are then compared to a mathematical model that
attempts to quantify all the factors that impact the travel time of
the electromagnetic waves on their way from the pulsar to the
Earth. In practice, a number of complications need to be dealt with, as outlined below. A more advanced
approach is to use multiple pulsars -- an ``array'' of
pulsars -- to look for signals that correlate between different pulsar
pairs. Such experiments are called ``pulsar timing arrays''
(\hl{PTAs}\index{PTA}).

\subsection{Template Profiles}

In its simplest form, pulsar timing could be based on the times when
the peaks in a train of pulses are detected. In order to increase the
measurement precision, one could also take the intensity-weighted
average arrival time of any given pulse. A far more powerful method,
however, is to use the information encoded in the \emph{shape} of the
pulse, to measure the arrival time relative to a standardised pulse
shape. This can be thought of as the ultimate, noise-free pulse
profile. Obtaining such a standardised pulse shape or
``\hbindex{template profile}'' is necessarily an iterative
process. Fundamentally, as many pulses should be averaged together as
possible. However, in order to align said pulses, an accurate pulsar
timing model should be used to predict the phase of subsequent pulses
to a precision far better than the time-resolution afforded by a
typical observation. To give an example, we aim to
predict arrival times to a precision of nanoseconds while typical
observations have phase resolution on the order of microseconds. The
use of all of the information contained in a complex pulse profile,
through the use of a template profile, allows one to achieve this
necessary magnitude of improvement.

Since the creation
of the template profile itself is the very start of the path towards a
functional pulsar timing model, typically the entire process gets
iterated so that the template profile and the timing model can both
improve until their solutions converge.

A danger in the creation of template profiles is so-called
``self-standarding''. This is a phenomenon that occurs when the data
that are being used in the timing, are also used to construct the
template profile. Specifically, if the number of observations that are
combined to construct the template profile is too low (a typical rule
of thumb is that ``too low'' is less than a few thousand), then the
noise within the observation can be ``recognised'' in the template
profile, leading to inaccurate offset measurements \citep[as
  illustrated clearly in Appendix A1 of][]{hbo05a}. Consequently, many
timing experiments make use of analytic models to describe the
template profile. These may not always be able to perfectly model all
the pulse-shape features, but they avoid timing corruptions caused
by self-standarding. Alternatively, a smoothing filter may be applied
to the template, in order to reduce the correlating noise 
\citep{dfg+13}. 

It was mentioned earlier that pulse profile shapes typically change
with the observing frequency. This should ideally be taken into
account when constructing the template profile, i.e.\ the template
profile should effectively have a dependence on observing frequency,
too. In most published works this has not been the case because the
bandwidth of pulsar observations used to be sufficiently narrow that
frequency evolution of the pulse profiles was effectively
undetectable, but over the last decade (fractional) bandwidths of
observing systems have increased so significantly that so-called
``frequency-resolved'' template profiles are rapidly becoming the norm
rather than the exception. Also in this case, one can either create an
analytic description of the profile in two dimensions 
\citep{pdr14,ldc+14} or use a template purely based on accumulated
data \citep{dvt+19}.

\subsection{Template Matching}

Once a template profile has been created, it can be used to calculate
the ToAs of the various observed pulse trains. Since most pulsars are
so faint that individual pulses cannot be detected and because single
pulses are usually not all alike, standard pulsar-timing experiments
do not time individual pulses, but average subsequent pulses modulo
the pulse period\footnote{At this point the question of which pulse is
being timed exactly, is in principle an arbitrary choice, but the most
commonly used pulsar-timing software uses a pulse in the centre of the
observation.}. This averaging procedure is commonly referred to as
\hbindex{folding} -- it reduces the time resolution of the observations
while \emph{phase} resolution is maintained. Typical values for the
time resolution of pulsar-timing data after folding, are anywhere from
minutes to one hour, depending on the brightness of the pulsar and the
goal of the experiment. Phase resolution is defined by the number of
bins across the profile, with typical values ranging from 128 to 2048,
depending on the sensitivity of the telescope and the bluntness or
sharpness of the pulse profile.

The folded pulse profiles could be cross-correlated with the template
profile in order to achieve the phase of the observation -- which can
then be added to the observation's time stamp in order to achieve a
ToA. In practice this measurement is commonly undertaken in the
Fourier domain, as explained in detail in the appendix of 
\citet{tay92}. Since an offset in the cross-correlation is equivalent
to a phase gradient in the cross-power spectrum, typically ToAs are
determined by least-squares fitting the phase gradient in the
cross-power spectrum of the template profile and the folded
observation. The phase offset resulting from this is then added to the
observation's time stamp.

The measurement uncertainty of these phase offsets -- and hence of the
ToAs -- is an important value as well, since many pulsars appear to
have highly variable flux densities (a process caused by the
interstellar medium, called scintillation), which means that not every
ToA carries as much information and hence should not be weighted
equally. Specifically, the pulse profile that is to be timed will
contain a certain amount of white noise called \hbindex{radiometer
  noise}, which depends on the observational properties of the pulsar
and the observing system as follows \citep{lk05}:
\begin{equation}\label{eq:RN}
  \sigma_{\rm ToA} \propto S/N = \beta\sqrt{n_{\rm p}t_{\rm int}\Delta
    f}\frac{T_{\rm peak}}{T_{\rm
      sys}}\frac{\sqrt{W\left(P-W\right)}}{P},
  \end{equation}
where $\sigma_{\rm ToA}$ is the ToA uncertainty, S/N is the
signal-to-noise ratio of the observation, $\beta$ is a factor which
describes instrumental losses, e.g.\ due to digitisation, $n_{\rm p}$,
$t_{\rm int}$ and $\Delta f$ are respectively the number of
polarisations combined, the integration time and the bandwidth of the
observation, $T_{\rm peak}$ is the brightness temperature of the
pulsar at the peak of its profile and $T_{\rm sys}$ is the brightness
temperature (i.e.\ noise) of the observing system, which typically
includes corrections for cable losses, instrumental gain, spill-over
and sky noise, amongst other things. $W$ is the equivalent width of
the pulse profile, defined as the integrated pulse intensity divided
by the peak intensity and $P$ is the pulse period.

The radiometer noise is the most fundamental factor limiting
pulsar-timing precision, in the sense that it is present in all
observations and is determined to a large degree by the fixed
properties of the pulsar and the technical capabilities of the
telescope. Traditionally it was quantified as the formal uncertainty
of the phase-gradient fit described above, although it has been shown
that in the low-S/N regime this can cause irregularities 
\citep[see][App.\ B]{abb+15}, leading people to either remove ToAs
below a certain S/N level (e.g.\ requiring S/N $> 8$) or to apply more
advanced, Monte-Carlo-based uncertainty estimations, as was proposed
as ``good pulsar timing practice'' by \citet{vlh+16}. 

\subsection{Timing Model Determination}

Once the ToAs have been measured, they need to be compared to
predicted arrival times provided by a pulsar timing model. A
\hbindex{timing model} is a mathematical formula that predicts the arrival
time of a pulse based on a set of timing-model parameters. Generally,
the timing model is defined in two steps \citep[see
  also][]{tay92,ehm06}. In the first step, the measured pulse arrival
time $t_{\rm obs}$ is referred to a time of emission at the pulsar
$t_{\rm PSR}$. This is achieved by accounting for all known
propagation and geometric delays:
\[
t_{\rm PSR} = t_{\rm obs} - \Delta_{\odot} - \Delta_{\rm IISM} -
\Delta_{\rm Bin}.
\]
Specifically, first the pulse ToA is
transferred to the Solar System barycentre (i.e.\ corrected by a delay
$\Delta_{\odot}$), which is the inertial
reference frame most commonly used in pulsar timing. This
transformation includes correction factors for relativistic effects
caused by the mass distribution in the Solar system and the Earth's
orbital and rotational velocity, for atmospheric propagation delays
(note these have mostly been neglected to date but will become
important with the next generation of radio telescopes), for
light-propagation times (the so-called Roemer delay), parallax
effects, frequency-dependent propagation delays induced by the
Solar Wind and corrections for the observatory clock. 

After the ToAs have been transferred to the Solar System barycentre,
interstellar propagation delays ($\Delta_{\rm IISM}$) are corrected
for. As described in more detail in the next section, these delays
have long been treated as dependent on the observing frequency, but
constant in time. With increased measurement precision provided by
wider bandwidths and lower observing frequencies, time-variable models
of interstellar dispersive delays are now becoming more common.

For pulsars that inhabit binary systems, there is one further
transformation, namely from the barycentre of the binary system to the
pulsar (delay $\Delta_{\rm Bin}$). This includes the Roemer delay
based on a Keplerian description of the binary orbit, but can also
contain relativistic effects such as the Shapiro Delay (time dilation
caused by the companion star's gravitational field), the Einstein
delay (time dilation caused by the pulsar's gravitational field and
gravitational redshift) or a host of other more complex effects,
dependent on the binary's properties. \citep[See][for a complete
  listing.]{ehm06}

Once the time of emission $t_{\rm PSR}$ is determined, it can be
converted to a rotational phase based on a spin-down model that is
usually simply described as a Taylor expansion:
\begin{equation}\label{eq:phase}
\phi\left(t_{\rm PSR}\right) = \nu \left( t_{\rm PSR} - t_0 \right)
+\frac{1}{2}\dot{\nu} \left( t_{\rm PSR} - t_0 \right)^2 +\ldots, 
\end{equation}
where $\nu$ is the spin frequency of the pulsar, $\dot{\nu}$ its first
derivative and $t_0$ an arbitrary reference epoch. Standard
electromagnetic theory predicts a second frequency derivative
$\ddot{\nu} = \frac{3\dot{\nu}^2}{\nu}$, but in practice this is
immeasurably small in the case of MSPs and is typically obscured by
other effects (so-called timing noise, see further) in most slow
pulsars. Consequently, by default pulsar timing models contain a spin
frequency and frequency derivative but not usually any higher-order
spin frequency derivatives.

Initial timing models are derived from pulsar-search
observations. These are raw time series that are not folded, but
instead are Fourier transformed in order to obtain an instantaneous
pulse period. By monitoring the pulse frequency evolve over several
such observations, an initial estimate of the spindown and of the
binary orbit can be determined. This then
constitutes an initial timing model that can be used to predict the
pulse frequency for future observations, allowing the data to be
folded in real-time, which makes observations much less demanding in terms
of data-storage and processing power requirements.

When a basic timing model has been constructed that is able to predict
the arrival time of future observed pulses to well within a pulse
period, it is said that ``\hbindex{phase connection}'' has been
achieved. From this point forward, the phases calculated in
Equation~\ref{eq:phase} can be used to improve the timing
model. Effectively, these phases should all be zero if the timing
model was perfect and no corrupting noise sources were
present. Consequently, any deviation from zero highlights effects
which do affect the observations, but are not taken into account
(correctly) by the timing model. These differences between the
observation and the model are called the \hbindex{timing residuals} and
they lie at the core of pulsar-timing analyses. Indeed, the art of
pulsar timing is to optimise and extend the timing model to decrease
these timing residuals. In the optimal case, the timing residuals will
only consist of white noise which is accurately quantified by the
uncertainties of the ToAs. In this case the timing is fully dominated
by the \hbindex{radiometer noise} described earlier.

\begin{figure}[]
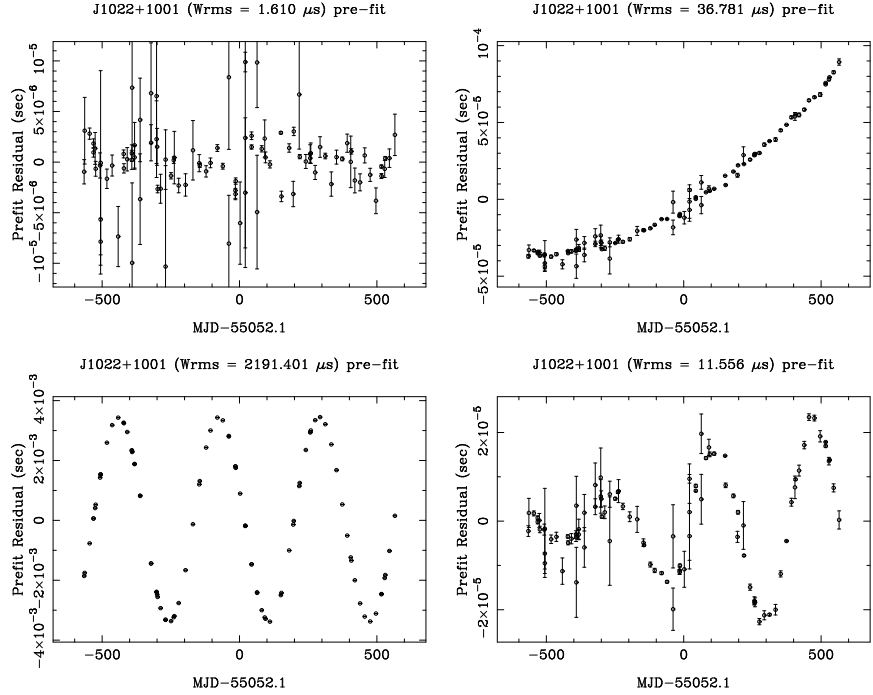

  \includegraphics[height=0.50\textwidth,angle=-90]{actual.ps}
  \includegraphics[height=0.50\textwidth,angle=-90]{F1.ps}
  \includegraphics[height=0.50\textwidth,angle=-90]{RADEC.ps}
  \includegraphics[height=0.50\textwidth,angle=-90]{PMRA.ps}
  \caption{Examples of pulsar timing residuals. The top-left figure
    shows a typical PTA data set taken from \citep{mhb+13}. The
    residuals are not purely white in this case, most likely due to a
    combination of uncorrected variations in interstellar propagation
    delays and interstellar dispersion. In the top-right figure the
    impact of a 1\% change in the spindown is demonstrated, leading to
    a clear quadratic trend in the residuals as the spin period gets
    increasingly incorrect as time progresses. The bottom-left figure
    shows a positional offset of 0.1 arcsec in both right ascension
    and declination, leading to an annual sine wave with constant
    amplitude. The bottom-right plot shows what happens in contrast if
    the position is correct (at the reference epoch near the start of
    this data set) but the proper motion is 10\% incorrect. This
    causes a linear increase in positional error and hence induces an annual
    sine wave with linearly growing amplitude.}
  \label{fig:signatures}
\end{figure}

Each parameter that is contained in the timing model affects the
timing residuals in a well-defined way, which is called the
\hbindex{timing signature} of this parameter. Consequently, simply by
visual inspection of the timing residuals, particular errors can
sometimes easily be picked out, as shown in
Figure~\ref{fig:signatures}. In most practical cases the uneven
spacing between observations, the variability in the ToA uncertainty
and the combination of multiple timing signatures in incomplete or
outdated timing models make the analysis of timing residuals rather
more complex than these simplified examples suggest.
 

\subsection{Interstellar Propagation Delays}

A particular source of difficulty when analysing pulsar-timing data,
is the impact of the ionised part of the interstellar medium (also
referred to as the \hbindex{IISM}). The refractive index of the
interstellar medium is determined by the plasma frequency, which is
dependent on the local electron density \citep{lk05}:
\[
f_{\rm p} = \sqrt{\frac{e^2n_{\rm e}}{\pi m_{\rm e}}} \approx
8.5\,{\rm kHz}\sqrt{n_{\rm e}},
\]
where $e$ is the electron charge, $n_{\rm e}$ is the electron density
in units of cm$^{-3}$ and $m_{\rm e}$ is the electron mass. Given that
plasma frequency, the refractive index for a photon with frequency $f$
can be determined as: $\mu = \sqrt{1-\left(\frac{f_{\rm
      p}}{f}\right)^2}$. 
Since the group velocity of electromagnetic waves is dependent on the
refractive index ($v_{\rm g} = \mu c_0$), this leads to a
frequency-dependent group velocity which is observable as a
frequency-dependent propagation delay:
\[
\Delta t = \frac{f_1^{-2}-f_2^{-2}}{K}{\rm DM},
\]
where \hl{DM}\index{dm@DM|see {dispersion measure}} is the
``\hbindex{dispersion measure}'' defined below and the constant $K =
\frac{1}{\mathcal{D}} = 2.41\times 10^{-4}$\,MHz$^{-2}$pc/cm$^{-3}$/s
is the inverse of the ``dispersion constant''. Theoretically, the
dispersion constant could be determined to higher precision
(specifically, $\mathcal{D} \equiv \frac{e^2}{2\pi m_{\rm e}c}$), but
in pulsar timing it has traditionally been fixed as given above 
\citep{kul20}.

The dispersion measure is straightforwardly defined as the integrated
electron content along the line of sight between the telescope and the
pulsar:
\[
{\rm DM} = \int_0^D n_{\rm e} {\rm d}l,
\]
where $n_{\rm e}$ is the electron density in cm$^{-3}$, $D$ is the
pulsar distance in pc and DM is typically expressed in units of
pc/cm$^{-3}$. However, given that in most pulsar-timing software
$\mathcal{D}$ has been defined fixed at the above-mentioned value and
since the actual observable is $\mathcal{D}\times$DM, in practice most
DM measurements would require a slight correction before being
interpreted as physical electron density measures, as described by 
\citet{kul20}. 

Due to the typically high spatial velocities of radio pulsars 
\citep[up to 1000\,km/s and beyond][]{cvb+05}, the line of sight along
which the pulses travel to Earth sweeps through interstellar space;
and due to the numerous turbulent structures that are present
throughout that space \citep{ars95}, this motion causes the DM to
change as a function of time. While such variability has long been
known to exist, it was mostly detectable as a slowly-varying,
red-noise process. Since the turn of the century, however, with
improved instrumental sensitivity, ever more precise measurements of
DM variations in time have been detected and accurate modelling of
DM(t) is becoming a complex and essential part of pulsar timing
experiments \citep{jml+17,kcs+13}.

One particular component that contributes to temporal variations in
DM, is the \hbindex{Solar Wind}. Due to the annual motion of lines of sight
-- particularly for pulsars near the ecliptic plane -- the additional
dispersion caused by the Solar Wind has a clear annual signature,
which is typically modelled straightforwardly by assuming the Solar
Wind to be homogeneous and spherically symmetric 
\citep{ehm06}. Generally, it is assumed that such a straightforward
model would suffice for the purposes of high-precision pulsar timing,
except perhaps closest to the Sun, so in addition to the spherical
models, PTAs have tended to remove ToAs for observations that took
place within 5-10 degrees of the Sun \citep{vlh+16}. It has been
attempted to extrapolate optical observations of the Sun to derive a
more detailed, inhomogeneous model of the Solar Wind for pulsar-timing
purposes \citep{yhc+07b}, but while this model was shown to provide an
accurate spectrum of inhomogeneities, it does not provide accurate
corrections for pulsar-timing experiments \citep{tvs+19}. 

In addition to dispersion, the IISM introduces a host of other
propagation effects, as recently reviewed by \citet{sti13}. In most
cases these effects are not limiting timing precision yet, although
time-variable scattering (also referred to as multi-path propagation
-- a phenomenon that widens the pulse shape through increased travel
path lengths) has been shown to be relevant in the timing of at least
one MSP \citep{lkd+17b}. 

\subsection{Timing Noise}

Probably the hardest effect to mitigate in pulsar timing is the
so-called ``\hbindex{timing noise}''. This term is generically applied to
any timing residuals that are not white noise and cannot be corrected
for by any of the deterministic timing-model parameters or by
frequency-dependent DM models. Presumably this typically
long-term noise is caused by inherent rotational instabilities in the
neutron star itself \citep{kop97b}, although the physical mechanism is
as yet not known.

Timing noise has been studied extensively in slow pulsars 
\citep{lhk+10,hlk10}, where it is highly common. In MSPs, timing noise
has been shown to be far less common, or to exist only at much
lower levels \citep{vbc+09,lsc+16}. Nevertheless, as the length of
pulsar-timing data sets grows and the timing precision increases, the
prevalence of timing noise -- and the importance of mitigation
techniques -- continues to increase even in MSP timing projects 
\citep{abb+18}.

\subsection{Other Noise Sources}

After the timing model is optimised and the IISM variations are
modelled and corrected for, ideally the timing residuals should be
spectrally white and normally distributed. In practice a wide range
of effects can negatively affect the timing, as recently reviewed by 
\citet{vs18}, although in practice the primary impact aside from the
IISM and timing noise, is pulse phase \hl{jitter}\index{jitter noise},
also known as \hl{SWIMS}\index{swims@SWIMS|see {jitter noise}}. The work
by \citet{lcc+16} on 37 MSPs, shows that jitter is relevant primarily
at low frequencies (particularly at observing frequencies below 1 GHz)
but less so at higher frequencies, where high-precision pulsar timing
is most commonly done. Since the importance of pulse jitter is
strongly dependent on the sensitivity of the telescope (and on the
available bandwidth), jitter will become an even more important source
of noise in the next generation of radio telescopes 
\citep{lkl+12}. One approach to avoid jitter-dominated timing, would
be to divide up large interferometric telescopes into less-sensitive
sub-arrays which allows more effective scheduling with lower jitter
noise \citep{swk+20}. Another approach is to mitigate jitter noise in
post processing as demonstrated in \citet{ovh+11,ovdb13}.

\subsection{Pulsar Timing Software}

All of the analysis described above tends to be carried out with two
separate types of data-analysis tools. Firstly one needs software
that can process the raw observation files: the so-called archives
that store radio-wave intensity as a function of polarisation, pulse
phase, frequency and time. Secondly one needs model-fitting software
that analyses the ToAs and the related timing models.

The primary software package that is used globally to analyse pulsar
archives in the context of pulsar timing, is
\textsc{\hbindex{psrchive}}
\citep{hvm04,vdo12}\footnote{\url{http://psrchive.sourceforge.net}}. The
only exception to the use of \textsc{psrchive} for the analysis of
pulsar archives and the creation of ToAs is the possible creation of
broadband ToAs with analytic, frequency-dependent templates. For this
purpose, the purpose-built \textsc{\hbindex{pulseportraiture}}
software
\citep{pdr14}\footnote{\url{https://github.com/pennucci/PulsePortraiture}}
is used increasingly commonly.  For the analysis of ToAs and timing
models, however, a larger variety of software packages has been
developed. The most common ToA-analysis package is
\textsc{\hbindex{tempo2}}
\citep{hem06}\footnote{\url{https://bitbucket.org/psrsoft/tempo2/}},
which is fundamentally a C/C++ translation of the much older, Fortran
77-based \textsc{tempo} package 
\citep{nds+15}\footnote{\url{https://github.com/nanograv/tempo}},
which is still being used, often in parallel with
\textsc{tempo2}. More recently, the \textsc{pint} package was
developed 
\citep{lrd+19}\footnote{\url{https://github.com/nanograv/PINT}}, which
is an independent timing package written in Python and which is
primarily used in North America.

Extending pulsar-timing software to constrain or detect correlated
signals, such as those from GWs, is a non-trivial effort. Whereas
frequentist methods have occasionally been implemented as part of
packages like \textsc{tempo2}, it has become far more common to build
independent software packages that are specifically aimed at Bayesian
analyses of the timing model -- including correlated signals. These
packages tend to be written in Python and use Python wrappers around
the source code of the standard pulsar-timing packages mentioned above
-- most commonly \textsc{tempo2}. The most recently used package for
such advanced ToA analysis (including GW analyses) is
\textsc{enterprise} 
\citep{evtb19}\footnote{\url{https://github.com/nanograv/enterprise}};
two commonly used earlier packages are \textsc{temponest} 
\citep{lah+14}\footnote{\url{https://github.com/LindleyLentati/TempoNest}}
and \textsc{piccard} 
\citep{van16}\footnote{\url{https://github.com/vhaasteren/piccard}}.

\subsection{Gravitational Waves and Other Correlated Signals}

In the previous sections we focussed on pulsar-timing effects and
phenomena that affect each pulsar independently in fully unrelated
ways. This is typical of the traditional single-pulsar timing
experiments that have been the mainstream of pulsar-timing research
ever since the first pulsar discovery. However, thoughout the 1980s
the realisation developed that some signals might have timing
signatures that correlate between pairs of pulsars 
\citep{hd83,rom89,fb90}. These signals would require a new, more
complex analysis as they require a joint analysis of a larger number,
say an array, of pulsars. The concept of the PTA was born, but would
not come to full fruition until the start of the new millennium, after
the number of known pulsars (and the number of PTA-worthy pulsars) had
dramatically increased following a couple of highly successful
surveys, primarily in the inner Galaxy \citep{mld+96,mlc+01}. In the
following paragraphs, an overview will be given of typical correlated
signals in pulsar-timing data, with particular focus on the signature
of gravitational waves. 

\subsubsection{Correlated Signals in Pulsar Timing Data}\label{sec:correlated_signals}

As described above, pulsar timing is effectively a model-fitting
exercise where deterministic parameters get optimised as increasing
amounts of data constrain the timing model. In addition to the
deterministic components of the timing model, there are
non-deterministic effects like DM variations or jitter noise that
affect the timing differently for each pulsar. Finally, there are
signals -- both deterministic and non-deterministic -- that affect all
pulsars in similar ways. Three types of such correlated signals have
been described to date \citep{rom89,fb90}:
\begin{itemize}
  \item a monopolar signal that would most
    likely arise from imperfections in the reference clock 
    \citep[e.g.]{hgc+20},
  \item a dipolar signal that is most likely related to the
    Solar-System ephemerides \citep[e.g.]{gllc19} and
  \item a quadrupolar signal that is predicted to arise from
    gravitational waves \citep{hd83}.
\end{itemize}
These correlated signals require more complex approaches: Solar-System
ephemeris models could be updated with the pulsar-timing data 
\citep{chm+10,gllc19,vts+20}, but clock signals and gravitational-wave
signatures are in essence random and hence the actual correlations
between timing of different pulsars need to be used in order to
determine those. For monopolar (or clock) signals, this has been
successfully achieved a number of times \citep{rod08,hgc+20}, but
quadrupolar (or gravitational wave) signal have so far not been
unambiguously detected, partly also because all these signals
interact, making a clear detection of the highest-order correlated
signal (i.e.\ the gravitational waves) dependent on accurate
determination of all other correlated signals \citep{thk+16}.

\subsubsection{Effect of Gravitational Waves}

The effect of GWs on pulsar timing was first described by 
\citet{det79} and more recently clearly summarised by 
\citet{sv10b}. Fundamentally, a GW passing over the Earth-pulsar
system will introduce a time-variable redshift into the pulsed
signal:
\[
z\left(t,\hat{\Omega}\right) \equiv
\left(\nu(t,\hat{\Omega})-\nu_0\right)/\nu_0,
\]
where $\Omega$ is the
direction of propagation of the GW, $\nu_0$ is a reference frequency
and $\nu(t,\hat{\Omega})$ is the observed pulse frequency which is
defined by the geometry of the system (pulsar and GW position with
respect to the observer) and the GW properties (polarisation and
amplitude). The integral of these redshifts quantifies the impact on the
timing residual:
\[
r(t) = \int^t_0z(t^{\prime},\hat{\Omega}){\rm
  d}t^{\prime}
\]
for an observation taken a time $t$ after the first observation in our
data set. Furthermore, the observed redshift is only dependent on the
perturbation of the space-time metric at the position of the pulsar at
the time when the pulse was emitted; and the space-time perturbation
at the location of Earth when the pulse was received.

\begin{figure}[]
  \includegraphics[width=\textwidth]{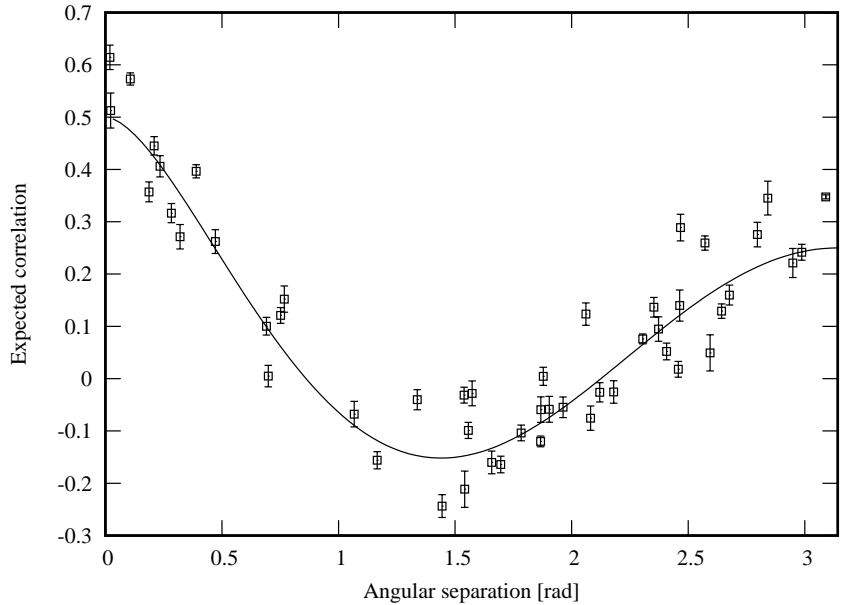}
  \caption{Correlated impact of a stochastic background of GWs on
    pulsar-timing residuals. Shown is the so-called Hellings-and-Downs
    curve \citep[after][]{hd83} as solid black line, which quantifies
    the correlation induced into pulsar-timing residuals of sets a
    pulsars as a function of the angular separation between those
    pulsars on the sky. For a single GW the shape would be similar,
    but would be dependent on the orientation of the pulsar pair with
    respect to the GW source; for a stochastic background only the angle
    between the pulsars matters. Square points show simulated
    measurements for an optimistic realisation of a PTA experiment.}
  \label{fig:HD}
\end{figure}

A more straightforward way of putting this, is that the GW impact on
the pulsar timing residuals has two equally large components: a
so-called ``\hbindex{pulsar term}'' that quantifies the impact of the
GW on the emission of the pulse; and an ``\hbindex{Earth term}'' which
quantifies the impact of the GW on the detection of the pulse on
Earth. For non-evolving, sinusoidal GWs, the Earth and pulsar terms
will be identical except for a phase offset between the
two. Furthermore, the Earth term will affect all pulsars equally
(albeit modulated in a quadrupolar way, as described by the so-called
Hellings-and-Downs curve \citep{hd83}, see Figure~\ref{fig:HD}) whereas
the pulsar term will have a different phase offset for each pulsar,
since the phase of the pulsar term depends on the distance to the
pulsar.

For such mono-chromatic signals, this phase offset could in principle
be measured along with the pulsar distance, allowing extremely precise
localisation of both the GW source and the pulsars in the array 
\citep{lwk+11}. Such an experiment would, however, require a level of
timing precision that is not realistically achievable with present-day
telescopes (but may be achievable with the next generation Square
Kilometre Array or SKA). At present, therefore, the pulsar term is
typically considered a noise term, while the Earth term is the actual
quadrupolar signal we hope to detect. The amplitude of the correlated
signal will give insights into the GW's origins (see the next
Section), whereas the shape of the correlation curve could constrain
fundamental physics of gravitational waves, like their polarisation
properties \citep{ljp08} and propagation speed
\citep{ljp+10}. Furthermore, deviations from the theoretically
expected Hellings-and-Downs curve can be expected for anisotropic
backgrounds of GWs, where a few bright sources stand out above the
background and cause a correlation function that is not only dependent
on the angle between pulsars, but also on their location on the sky
\citep{tmg+15}. 

%
%

\section{GW Sources in the PTA Band}
\label{sec:3}

The sensitivity of PTAs to GWs is limited in terms of the GW frequency
by the length of the observing time span and the cadence of the
observations. Specifically, PTAs are most sensitive near a frequency
of $1/T$ where $T$ is the length of the data set, i.e.\ on the order
of a decade or more, which corresponds to a frequency on the order of
nanohertz. Since the GW impact is a change in the pulse frequency
\citep{det79}, but the timing residuals are phases, i.e.\ effectively
integrals over pulse frequency, the sensitivity of PTAs decreases with
increasing frequency down to the Nyquist frequency $2/C$ where $C$ is
the cadence of the observations, typically of the order of weeks or a
month, which corresponds to a frequency cut-off of the order of
microhertz.

\begin{figure}
  \includegraphics[width=\textwidth]{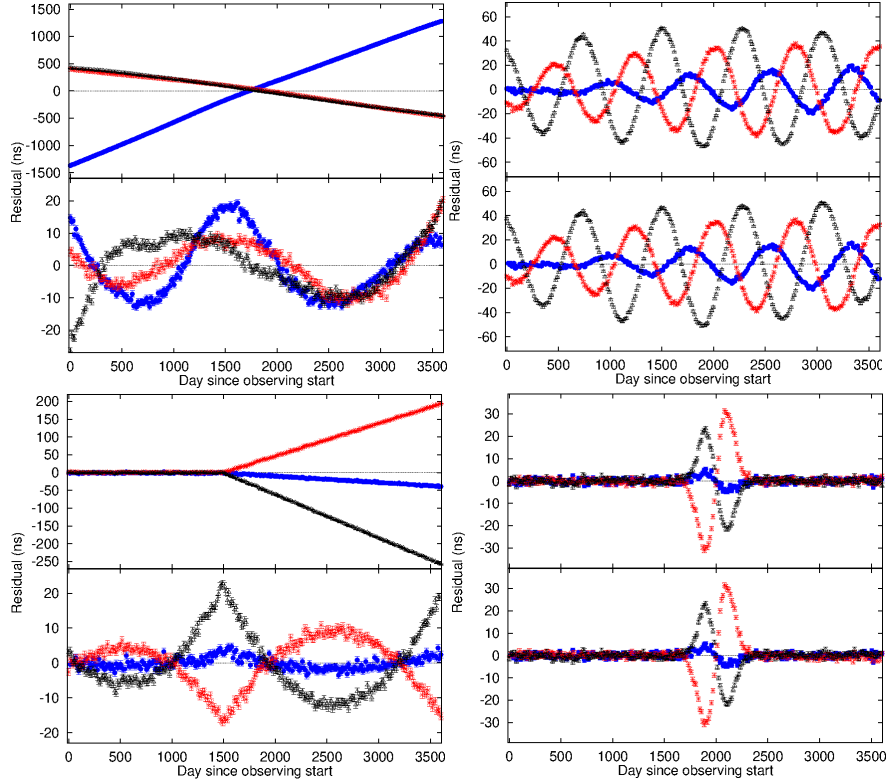}
  \caption{Example timing residuals for four GW types on three
    different pulsars. These four panels show what the timing
    residuals that are caused by GWs could look like. The top-left
    panel shows the influence of a GWB (with characteristic strain
    amplitude $10^{-15}$ and spectral index $-2/3$), the top-right
    panel that of a CW (from a $10^9$\,$M_{\odot}$ equal-mass binary
    supermassive-black hole at redshift $z=0.01$), the bottom-left
    panel shows a BWM (with strain amplitude $5\times 10^{-15}$) and
    the bottom-right panel shows a GW burst (without memory and with
    arbitrary waveform). The three different colours show the impact
    on different pulsars (i.e.\ different sky locations), with red
    showing the timing residuals of PSR~J0437$-$4715, blue those of
    PSR~J1012+5307 and black those of PSR~J1713+0747. The simulated
    measurement uncertainty is 1\,ns and no intrinsic spin noise or DM
    variations were included in the simulations. The top figure
    in each panel shows the pre-fit residuals (i.e.\ the raw GW
    signature), whereas the bottom plot shows the post-fit
    residuals. The difference between these is caused by fitting of
    the standard timing-model parameters. As can be seen, most of the
    power of the GWB is absorbed in the timing-model fit, since its
    long-term signature resembles the timing signatures of the pulse
    period and spindown, which absorb most of the signal. Such
    absorption of GW residuals in common timing-model parameters is
    less likely to happen for CWs or bursts, unless they happen to
    have a periodicity that is close to a year (or an integer fraction
    or multiple thereof) or to the orbital period of the pulsar being
    timed.}
    \label{fig:classes}
\end{figure}

This frequency range makes PTAs particularly complementary with other
GW detectors like LIGO
\citep[10\,Hz--10\,kHz,][]{mha+16} and LISA
\citep[0.1\,mHz--1\,Hz,][]{aab+17}; and competitive with proposed GW
detection methods based on space-based VLBI \citep{bkpn90,bf11}. The
difference in GW frequencies furthermore implies that different
sources of GWs can be expected to be detectable with
PTAs. Specifically, four types of sources are anticipated, as
described below. Example timing residuals induced by these four types
of GW are shown in Figure~\ref{fig:classes}. 

\begin{description}
  \item[Gravitational wave background (\hbindex{GWB}):] A GWB is a
    superposition of GWs from a large number of GW sources that add
    incoherently. The most likely GWB in the PTA frequency range is
    widely expected to arise from \hbindex{supermassive black hole} binaries
    and predictions for its spectrum and amplitude are based on
    simulations such as the Millennium Run \citep{av10} or the
    Illustris simulation \citep{kbh+17}. Specifically, the power
    spectrum of this GWB is expected to have a power-law shape with
    slope $-2/3$ and is likely to flatten or even tip over at
    GW frequencies lower than $\sim0.1$\,yr$^{-1}$
    \citep{csc19}. Alternative backgrounds have been proposed
    \citep[see, e.g.][]{bb08,sch15,sbs12}, with differences in both
    spectral shapes and amplitudes, implying that an eventual
    detection would be able to differentiate between the origins of
    the GWB, or would be able to place constraints on the
    galaxy-merger history of the Universe.
  \item[Continuous waves (CWs):] Continuous GWs in the PTA band are
    expected from single supermassive black-hole binaries that are
    close enough to Earth to stand out beyond the GWB. 
    With improved sensitivity, a detection of CWs is generally expected to follow within a few years from a GWB detection \citep{rsg15,mls+17}.
  \item[\hbindex{GW bursts}:] Single GW events like close encounters
    between supermassive black holes or some \hbindex{cosmic string}
    interactions \citep{dv01}, could result in a single burst of GWs,
    which might be detectable provided the burst itself lasts
    sufficiently long for it to affect multiple subsequent pulsar
    observations (i.e.\ at the very least days long); and provided the
    burst is sufficiently bright to stand out of the noise.
  \item[Bursts with memory (\hbindex{BWMs}):] Bursts that are too
    faint to be detected directly, could still be detected as a
    ``memory event'' or a BWM. In this case it is the permanent
    deformation of the space-time metric \citep{fav09} that has a
    lasting impact on the pulse frequency, causing the impact on the
    timing to accumulate over time.
\end{description}

\section{Present PTA Constraints}

Around the world, collaborations have emerged to carry out the
large-scale observational campaigns required for GW detection with
PTAs. The Australian Parkes PTA \citep[\hbindex{PPTA}, see][]{mhb+13}
was the first one, commencing observations in 2005 and revising some
of the original theoretical work \citep{jhlm05,jhv+06}. Centred around
the 64-m Parkes radio telescope, they have been monitoring between 20
and 30 pulsars using both dedicated observations \citep{krh+20} and
archival data \citep{vbc+09}. The PPTA last placed a limit on the GWB
in 2015 \citep{srl+15}, finding that the normalised amplitude at a
frequency of 1\,yr$^{-1}$ must be lower than $10^{-15}$ with 95\%
confidence -- which started to get into the area of theoretically
predicted amplitudes at the time; this limit is still the most
constraining bound on a GWB in the PTA band to date. Most recently,
the PPTA has focussed its efforts on alternative sources of GWs, such
as ultra-light scalar-field \hbindex{dark matter} \citep{pzl+18} as well as CWs
and BWMs \citep{mzh+16}; alongside a more intensive commitment to the
global International PTA (see below) and instrumental development
\citep{hmd+20}.

The European PTA \citep[\hbindex{EPTA}, see][]{dcl+16} also commenced
in 2005, soon after the PPTA and also relies on a combination of
specific PTA data and archival monitoring data, adding up to a total
of 42 MSPs \citep{dcl+16}. To date, the EPTA has primarily used data
from the Jodrell Bank 76.2-m Lovell telescope, the 100-m Effelsberg
Radio Telescope, the 300-by-35-m decimetric radio telescope at
Nan\c cay observatory and the Westerbork Synthesis Radio Telescope
which is an interferometric array consisting of 14 antennae of 25-m
diameter.  Their most recent limit also dates back to 2015
\citep{ltm+15}, but is slightly less constraining, at $A_{\rm 1/yr} <
3.0\times 10^{-15}$. The same data set has been used to place
constraints on CWs \citep{bps+16} and possible anisotropies in the GWB
\citep{tmg+15}. Finally, dedicated, high-cadence, data were used to
place constraints in the microhertz regime \citep{psb+18}. In recent
years, the EPTA has primarily focussed on instrumental development
with new data-recording systems \citep[e.g.]{lkg+16}, the
commissioning of a new 64-m radio telescope in Italy
\citep[SRT, see][]{pmt+17} and the interferometric combination of all
mayor EPTA telescopes in the ``Large European Array for Pulsars''
(LEAP) project \citep{bjk+16}. The EPTA has also had a strong
involvement in the commissioning of the LOw-Frequency ARray
\citep[LOFAR, see][]{vwg+13}, which has shown to be a useful telescope
for monitoring not only MSPs \citep{kvh+16} but particularly DMs
\citep[e.g.]{dvt+19}.

The third major PTA is the North-American Nanohertz Observatory for
Gravitational waves \citep[\hbindex{NANOGrav}][]{dfg+13}. NANOGrav has
been particularly involved in pulsar searches and as such their source
list has been continuously expanding, counting 48 MSPs at their
current data release \citep[\url{https://data.nanograv.org}
  and][]{lma+19}. The latest and most constraining limit on the
amplitude of the GWB, is $A_{\rm 1/yr} < 1.45\times 10^{-15}$ 
\citep{abb+18a}, slightly above the PPTA limit. Within the last few
years, NANOGrav has also placed significant bounds on BWMs 
\citep{aab+20a} and on CWs \citep{aab+20}; and placed a specific limit
on the proposed binary black-hole system in the radio galaxy 3C66B 
\citep{simt03,abb+20}.

The three original PTAs mentioned above joined forces to further
increase sensitivity and in an attempt to decrease the time to the
first GW detection in the nHz regime. As described by \citet{man13},
the first joint PTA meeting took place in 2008, but the formal
establishment of the International PTA (\hbindex{IPTA}) did not happen
until 2011. Since then, the IPTA has released two combined data sets 
\citep{vlh+16,pdd+19} and has indicated that an improvement in GWB
sensitivity by a factor of about two should be expected, but no full
GW analysis has been carried out on IPTA data so far, given the
complexities involved with the highly inhomogeneous nature of the data
set. These inhomogeneities and complexities \citep[discussed and
  listed in][]{vlh+16} are a specific challenge for any combined PTA
experiment and often requires additional research regarding detailed
aspects of the analysis, or development of new methods that are able
to deal with such inhomogeneous data. A lot of progress has been made
in recent years in this regard, particularly with the advent of
Bayesian analysis software packages like \textsc{temponest} 
\citep{lah+14,lsc+16} and \textsc{enterprise} \citep{evtb19}, amongst
others. So far, the first IPTA data combination has been used by 
\citet{hgc+20} to construct a pulsar-based time standard (thereby
solving for any monopolar correlations in the data), while 
\citet{cgl+18} used it to constrain errors in the Solar-System
ephemeris models used. A comprehensive GW analysis is planned for the
second data release. Meanwhile, analysis tools are being tested on
mock data challenges \citep{hml18,vs18,bbf+20}.

As a new generation of telescopes is being constructed on the pathway
to the SKA, a number of new PTAs have recently emerged. Specifically,
the Indian PTA \citep[\hbindex{InPTA},][]{jab+18} has been formed in
2018 and uses high-precision timing data from the upgraded Giant
Metre-wave Radio Telescope (uGMRT) and low-frequency data from the
Ooty Radio Telescope (ORT). The Chinese PTA 
\citep[\hbindex{CPTA}][]{lee16} uses the Five-hundred meter Aperture
Spherical Telescope (FAST), the Xingjiang Qitai 110-m Radio Telescope
(QTT) and a network of 100-m-class radio telescopes across China and
predict to go well below current sensitivity limits after even a few
years of observing. Finally, the South-African MeerKAT telescope 
\citep{jm16} is being used by the international \hbindex{MeerTIME}
consortium \citep{bbb+16} to produce (amongst other things) a highly
sensitive PTA data set, part of which will be taken at relatively high
radio frequencies, between 1.7 and 3.5\,GHz, thereby limiting the
impact of interstellar effects.

\begin{table}
  \caption{Summary of present limits on GWs from
    PTA data.}
  \label{tab:GWLimit} 
  \begin{tabular}{lrrrr}
    \hline\noalign{\smallskip}
    GW Type & EPTA & NANOGrav & PPTA & IPTA\\
    \noalign{\smallskip}\svhline\noalign{\smallskip}

    \multirow{2}{*}{GWB$^{a}$} & $3.0\times 10^{-15}$ &
    $1.45\times 10^{-15}$ & $1.0\times 10^{-15}$ & $1.7\times 10^{-15}$\\
                         & \citet{ltm+15}       & \citet{abb+18a}
    & \citet{srl+15} & \citet{vlh+16}$^{b}$ \\
\\
    
    \multirow{2}{*}{BWM$^{c}$} & -- & 1.5/yr & 0.75/yr & --\\

    & -- & \citet{abb+15b} & \citet{whc+15} & --\\

    \\
    \multirow{2}{*}{CW$^{d}$} & $1.5\times 10^{-14}$ &
    $3.0\times 10^{-14}$ & $1.7\times 10^{-14}$ & --\\
     & \citet{bps+16}$^{e}$ & \citet{abb+14} & \citet{zhw+14} & -- \\
    \noalign{\smallskip}\hline\noalign{\smallskip}

  \end{tabular}
  
  $^{a}$ Limits on the GWB are typically given as upper limits on the
  dimensionless strain amplitude at a frequency of 1/yr.\\
  $^{b}$ \citet{vlh+16} note that the limit derived from the IPTA
  data set was only indicative and not rigorous as a full analysis was
  deferred to a future paper.\\
  $^{c}$ BWMs can be quantified in many ways. In order to provide
  some comparative measure, this table presents the upper limit on the
  burst rate for bursts with normalised characteristic strain
  amplitude $10^{-13}$.\\
  $^{d}$ Limits for CWs are given as upper bounds on $h_0$, at a GW
  frequency of 10\,nHz. \\
  $^{e}$ \citet{bps+16} don't give a specific value; the number given
  is based on their Figure~3.
\end{table}

%
%
%
%

\section{Recent and Ongoing Improvements in PTA Sensitivity}

Much of the present interest in PTA experiments derives from the work
by \citet{jhlm05} who predicted that a GWB should be detectable after
a mere 5 years of timing on 20 pulsars, with a timing residual RMS of
100\,ns -- a level of precision that had only recently been
demonstrated to be achievable in practice 
\citep{vbb+01}. Subsequently, it became clear that the ideal PTA (20
pulsars, 5 years and 100\,ns RMS) was unlikely to ever become a
reality since the pulsar population is by nature highly inhomogeneous,
implying a few pulsars would be likely to be timed at better precision
than 100\,ns, but most probably would not. Consequently, scaling
relations were derived, first by \citet{jhv+06} and later by 
\citet{sejr13}, to allow fine-tuning of PTA experiments with a view to
optimising sensitivity and shortening detection time scales.

\citet{sejr13} showed that the S/N of a GWB in a PTA data set scales
with typical properties of the data set as follows:
\begin{equation}\label{eq:PTASens}
  S/N \propto N C^{3/26}A^{3/13}T^{1/2}\sigma^{-3/13},
\end{equation}
where $N$ is the number of pulsars, $C$ is the cadence of the
observations, $A$ is the amplitude of the GWB, $T$ is the length of
the timing data set and $\sigma$ is the RMS of the timing
residuals. While this analysis makes some basic simplifications in
terms of data homogeneity, it does show clearly the very strong
dependence of PTA sensitivity on the number of pulsars in the
array. Consequently, a large number of pulsar surveys have been
undertaken in recent years to increase the number of PTA-useable MSPs,
as described below. The dependence on all other parameters is far less
significant, except for the data length, which still adds
considerably. In the low-S/N regime, \citet{sejr13} show that a PTA's
sensitivity scales most strongly with the data length: $S/N \propto
T^{13/3}$, in close agreement with the earlier findings of 
\citet{jhlm05}. For long data sets the sensitivity is however affected
by the level at which other long-period signals can be mitigated. Most
significantly this refers to DM variations which can be mitigated
provided the observing set-up has been well chosen. Pulsar timing
noise is equally important, but in the absence of predictive models or
independent estimates of this noise source, the only possible approach
is to limit its impact by post-facto modelling and
subtraction. Finally, in the intermediate S/N regime (i.e.\ as we get
closer to detections rather than mere limits), the sensitivity to a
GWB is only weakly related to the timing precision of the MSPs, but
for single sources of GWs the timing precision is still the dominant
factor, consequently some efforts are being made to further improve
the timing precision of MSPs by instrumental improvements and building
of new telescopes. At the end of this section, a brief overview is
given about various studies that quantify how all of these
improvements are likely to affect the time to the first GW detection
with PTAs -- which mostly agree a detection within years to at most a
decade is highly likely. 

\subsection{Pulsar Surveys}

In order to increase the sample of MSPs that can be used in PTA
experiments, a number of pulsar surveys have been undertaken in recent
years and are being planned for the near
future. 
Specifically, two
long-lasting surveys have been running on the Arecibo radio telescope
for most of the past decade: the P-Alfa survey \citep{cfl+06} and the
Arecibo drift-scan survey \citep{dsm+13}. In addition, the Effelsberg
and Parkes radio telescopes (in Germany and Australia respectively)
are continuing their all-sky partner surveys HTRU North and South 
\citep{bck+13,kjv+10}. Parkes has simultaneously been equipped with
cutting-edge processing technology as part of the SUPERB 
\citep{kbj+18} survey, which aims to do real-time searches for pulsars
and fast radio bursts. Data from the Green Bank Telescope (GBT)
continues to be analysed in the Green Bank Northern Celestial Cap
(GBNCC) survey \citep{slr+14}. At low frequencies, the LOFAR telescope 
\citep{vwg+13} is finishing processing of the LOFAR Tied-Array All-Sky
Survey \citep[LOTAAS,][]{scb+19} and at the GMRT the GMRT
High-Resoluiton Southern Sky survey (GHRSS) is ongoing 
\citep{bcm+16}. New telescopes are also getting up to speed on pulsar
surveys, with the first successful discoveries published by the FAST
telescope \citep{yln13,qpl+19}, as part of the Commensal Radio
Astronomy FAST Survey (CRAFTS) and survey observations recently
commenced for the MeerKAT ``TRAPUM'' survey \citep{sk16}.


\subsection{IISM Studies}

In order to increase sensitivity of pulsar-timing data sets to
long-term signals like those expected from a GWB, it is of utmost
importance to understand, mitigate and model any long-term signals
that may be affecting the timing. Most such effects are deterministic
effects contained in the timing model, but two more complex sources of
red noise exist: timing noise and IISM noise. As discussed earlier,
the origin of timing noise has not been unequivocally determined and
consequently most models are rather ad-hoc power-law descriptions of
uncorrelated timing signatures. IISM noise (or DM variations) is
different, since it is the only known effect that causes a frequency
dependence in timing residuals.

The frequency dependence of IISM noise implies that in principle it
can be measured and modelled independently from all the other
timing-model parameters and correlated signals, because it can fully
rely on the frequency resolution of the data. Specifically, three
scenarios could be envisaged for measuring and correcting DM
variations in pulsar-timing data \citep{jhm+15}:
\begin{itemize}
  \item Multiple different observing bands: When more than one
    observing band is used, the frequency difference between the bands
    can be used as a lever arm that enables high-precision DM
    estimates. This idea has been implemented both with co-axial
    receivers and, more recently, with ultra-broadband receivers such
    as the UWL in Parkes \citep{hmd+20} and similar observing systems
    at the Effelsberg and Green Bank radio telescopes, often not using
    actual instantaneous observations, but by determining an average
    DM over a range of dates \citep{kcs+13}. In this scenario, care
    must be taken in the allocation of the observing time, since the
    DM sensitivity of the data will scale with the square of the
    observing wavelength, but the timing precision may not, depending
    on the spectral index of the pulsar. Specifically, since on
    average pulsars have a spectrum that is flatter than $\nu^{-2}$ 
    \citep{jvk+18} and since the sky background noise is steeper
    than typical pulsar spectra \citep{web74}, it is likely that the
    low-frequency band will still have superior sensitivity to DM, but
    have worse timing precision overall. For observations beyond
    1.4\,GHz, however, the situation often reverses, in that
    higher-frequency bands may have both less sensitivity to
    dispersion \emph{and} lower timing precision \citep[see,
      e.g.][]{lkg+16}. This implies a change in integration time
    depending on the observing frequency, may be in order. A more
    extensive analysis of post-correction ToA precision in this
    scenario, is given by \citet{lbj+14}.
  \item Low-frequency monitoring: As a possible way to mitigate the
    complexities of balancing DM and timing precision, one could
    attempt to monitor pulsars at low frequencies (generally at or
    below 400\,MHz) and derive \emph{independent} DM time series from
    those low-frequency data, to correct the higher-frequency
    data. This has the additional advantages that the DM modelling is
    now fully independent from the higher-frequency timing; and at low
    frequencies DM corrections could often be measured within a single
    observation, which avoids correlations with effects like timing
    noise or other timing-model parameters. This approach has its own
    drawbacks because other IISM effects like scattering also become
    more pronounced at low frequencies and so the DM measurements may
    be biased or corrupted. Finally, in some cases the differences in
    Fresnel scale\footnote{The Fresnel scale is a basic measure for
      the size of the Fresnel zone, which in turn is the region of
      space a signal can travel through between emitter and receiver.}
    at the top and bottom of the observing band make it possible that
    the actual space probed by electrons at different frequencies is
    slightly different -- and hence the DM as well. This results in a
    frequency-dependent DM, which has been theoretically predicted
    \citep{css16} and observed \citep{dvt+19}, but the overall impact
    of this phenomenon on PTA sensitivity has so far not been
    accurately quantified.
  \item High-frequency timing: Finally, with sufficiently sensitive
    telescopes, the main observing frequency could be moved to higher
    frequencies, where the IISM effects are weaker. Pulsars also tend
    to be fainter at those frequencies, but in the case of highly
    sensitive telescopes, this may be a blessing since it implies
    jitter noise will be less significant, as more pulses will need to
    be averaged per observation. However, with present telescopes,
    this approach may require too much observing time \citep{lkg+16}
    -- with more sensitive, future, telescopes it may become a
    realistic option.
\end{itemize}

When it comes to effects of the IISM, dispersion is only the peak of
the iceberg. Things get a lot more complex when we consider multi-path
propagation or \hbindex{scattering}. This phenomenon arises when some
of the photons get deviated from the straight line between the pulsar
and Earth due to refraction; and later get refracted back into the
line of sight. In its simplest form, for a thin scattering screen with
density inhomogeneities that follow a Kolmogorov spectrum, this should
cause a delay in photon arrival time which scales with the observing
frequency to the fourth power:
\[
\tau_{\rm scat} \propto \nu^{-\alpha},
\]
where the scattering index $\alpha$ is theoretically expected to be
4.4. In practice, \citet{bcc+04} have measured an average spectral
index of the scattering strength of $\alpha = 3.9\pm 0.2$, slightly
inconsistent with theory. More detailed studies at lower frequencies 
\citep{gkk+17} have shown that in many cases the frequency scaling of
the scattering was more consistent with a highly \emph{an}isotropic
scattering screen than with the typical Kolmogorov screen.

Regardless of the frequency scaling, the primary observable effect of
scattering in pulsar timing is that the pulse profile gets smeared out
and gets a characteristic exponential tail. This worsens timing
precision since it can wash out features and it may corrupt the DM
measurement (although absolute DM measurements may not be necessary
for pulsar-timing experiments anyway), but in principle this would not
affect GW sensitivity as long as the effect is constant in time.

\begin{figure}[]
  \includegraphics[scale=0.8,angle=0.0]{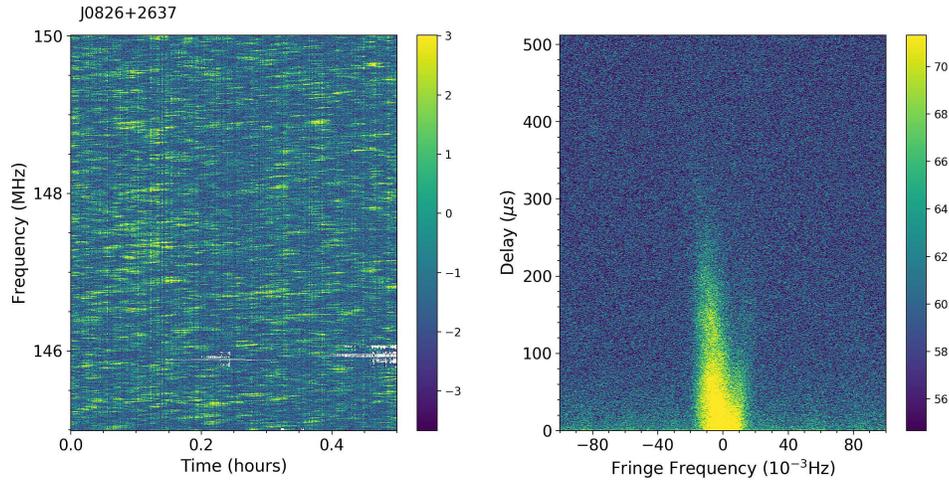}
  \caption{Dynamic and secondary spectrum of PSR~J0826+2637
    (PSR~B0823+26). Left: the dynamic spectrum of a half-hour
    observation on PSR~J08216+2637 with the LOFAR core. Shown is the
    pulsed intensity on an arbitrary colour scale (units are
    uncalibrated) for a
    segment of 5-MHz bandwidth, centred on 147.5\,MHz. The bright
    patches are called scintles and are caused by diffractive
    scintillation. Less clear is the higher-frequency corrugations
    that run diagonally across this dynamic spectrum and which are
    caused by a combination of diffractive and refractive
    scintillation. Right: this figure shows the secondary spectrum of
    the left-hand plot, i.e.\ the 2D Fourier transform, whereby the
    power levels are shown on a logarithmic scale. A highly asymmetric
    arc is visible, extending out to fractions of a millisecond at
    negative fringe frequencies (also called ``Doppler rates'') but
    only out to about 0.1\,ms at positive fringe frequencies. Figure
    courtesy of Ziwei Wu.}
  \label{fig:DS}
\end{figure}

The strength of scattering does change in time, though, as can most
readily be seen by inspecting a ``\hbindex{dynamic spectrum}'': a plot
of pulsed intensity as a function of frequency and time. Such
time-variable spectra often show changes in pulse intensity -- a
phenomenon known as \hbindex{scintillation}. Diffractive scintillation
(as shown in Figure~\ref{fig:DS}) occurs when photons travelling from
the pulsar to Earth meet refractive structures and get slight phase
shifts due to location-dependent refractive indices. These
phase-shifts cause constructive and destructive interference which are
seen as bright and dark pathes in the dynamic spectrum. Since
scattering and diffractive scintillation are effectively two different
observables caused by the same turbulent and diffractive structure, it
should not come as a surprise that they are related -- in fact, they
are inversely proportional \citep{lmj+16}. An extensive analysis of
PTA data carried out by \citet{lmj+16} quantified the variations in
diffractive scintillation and consequently estimated how variable
scattering is in typical PTA observations. This analysis showed that
at present levels of sensitivity, scattering variations are only
rarely a real concern \citep[with one exception studied in detail
  by][]{lkd+17b}, but in the next era of highly sensitive telescopes
(notably MeerKAT, FAST and SKA), this will probably change.

A further complication could arise from a higher-order effect where
scattering and diffractive scintillation combine in what are known as
``\hbindex{scintillation arcs}'' \citep{smc+01}. This primarily occurs
in highly anisotropic media, when the majority of the radiation comes
from the pulsar (essentially a point source), but significant amounts
of energy come from other, typically straight and narrow, structures
on the sky. It causes ripples across the dynamic spectrum, which are
more easily noticed in the 2D Fourier transform of the dynamic
spectrum -- this is also called the ``\hbindex{secondary spectrum}''
(see right-hand side of Figure~\ref{fig:DS}). The initial discovery of
secondary spectra is still relatively recent and since these tend to
be faint features which require high sensitivity and high resolution
(i.e.\ large data rates), their study has only developed
slowly. However, early studies by \citet{hs08} already showed how
these phenomena can impact timing significantly. Turning this around, 
\citet{rcb+20} showed how scintillation arcs can actually be useful
for timing, as they can provide independent constraints on
timing-model parameters. The study of how to use scintillation arcs,
or what effects they really do have on PTA experiments, has only just
begun, so at present their impact is not fully clear yet.

A final concerning occurrence that the IISM may create, are
frequency-dependent DMs (also often -- and confusingly -- named
``chromatic DMs''). The principle behind frequency-dependent DMs is
as follows: in wide-band observations, photons with a wide range in
wavelengths are observed. These photons did not all travel through the
same space -- in fact, since the Fresnel scale is
frequency-dependent, there is a bias that causes lower-frequency photons
to be able to travel through a wider region of space than
higher-frequency photons. If the variations in electron density are
sufficiently extreme -- or if measurement precision is sufficiently
high -- this would imply that the high-frequency photons may sample a
different electron distribution than the low-frequency
photons. Consequently, the DMs measured at the top and bottom of the
observing bands may be different because they refer to different parts
of space.

While the concept of frequency-dependent DMs had been known for much
longer, the theoretical description was first laid out by 
\citet{css16}. Detection of such a phenomenon is naturally complex
given the many other frequency-dependent effects described earlier,
but by looking at the time-difference of DMs measured at opposite
parts of a low-frequency observation, \citet{dvt+19} succeeded in
making the first clear detection of such chromatic effects. These
initial results showed rather a more complex picture than the theory
had predicted, clarifying that further research into
frequency-dependent DMs is required before a conclusive understanding
of their possible impact on PTAs can be drawn. Given the continuous
coverage over extremely wide frqeuency ranges of new telescopes like
the uGMRT, ngVLA, MeerKAT and the SKA, a much clearer understanding is
bound to arise within the next decade.

\subsection{Sensitivity Predictions}

As mentioned earlier, as PTAs edge closer to a GW detection, the
number of pulsars in the array is of key importance for the PTA's
sensitivity. Since predicting pulsar discoveries is infamously hard,
it is equally hard to make accurate predictions of PTA sensitivity
given future pulsar discoveries, yet several papers have demontrated
the validity of the adage ``More pulsars is more sensitivity''. Most
recently, \citet{kbh+17} showed that -- assuming ongoing regular
discoveries of MSPs that can be timed at high precision -- all PTAs
could hope to detect GWs within a few years; and all were virtually
guaranteed a detection within a decade.

Based on the most up-to-date predictions for a GWB in the PTA band and
realistic numbers from existing PTA experiments, \citet{rsg15} showed
that the most likely scenario would be that a GWB would be detected in
about one to two decades time. A detection of CWs was also not
unrealistic, but would probably take somewhat longer.  \citet{rsg15}
did not investigate the impact of increasing the number of MSPs in the
array, but did evaluate the impact of more sensitive systems -- the
SKA in particular and found that with the full SKA, a GWB should be
detectable within a few years; and CWs within about a decade.

The idea that a highly sensitive telescope could detect GWs within a
few years, even with existing pulsars, was also demonstrated by 
\citet{lee16}, who drew essentially the same conclusion for the
CPTA with its unprecedentedly sensitive set of telescopes. Also 
\citet{lkg+16} investigated the impact of improved sensitivity on PTA
detection time scales, this time in the context of improved receiver
and recording systems. Specifically, they estimated that the new
recording system at Effelsberg would improve the telescope's
sensitivity to a GWB by a factor of up to three compared to the status
quo, in only four years time. They furthermore continued the work by 
\citet{lbj+14} to demonstrate how wide-band and low-frequency
observing systems might aid the detection of GWs by efficiently
correcting DM variations and thereby keeping the timing RMS
low. \citet{vlh+16} approached the sensitivity improvements in a
very complementary way, showing that global collaboration and sharing
of data would lead to approximately a factor of two improvement in GW
sensitivity.

It is also possible that the coming years continue to bring non-detections; this
could occur if we encounter an unexpected instrumental noise floor, such as intrinsic
pulsar noise or a high level of ephemeris uncertainty. However, this
scenario is unlikely to be an issue. Pulsar noise can be overcome by
targeted noise modelling \citep[as in][]{hst+20}, or by simply adding
more pulsars to a PTA, which will beat down noise that is uncorrelated
between different pulsars, thus still raising our sensitivity to the
correlated GW signals. Regarding uncertainties in the Solar-System
ephemerides, as previously noted on page \pageref{sec:correlated_signals}, ephemerides uncertainties are correlated. The signal, however, is dipolar thus we can attempt to measure and remove it, even if there is some leakage between dipolar and quadrupolar signals \citep{thk+16}. In addition, it has been demonstrated that PTAs are already breaching the accuracy of published ephemerides, and techniques have been developed to overcome such uncertainties \citep{vts+20}.

Regardless, current upper limits on the GWB are \emph{already}
impacting our understanding of the evolution of galaxies and their
\hbindex{supermassive black hole} residents
\citep{srl+15,sb16,abb+18a,csc19}, in addition to placing novel
constraints on \index{cosmic string}cosmic strings \citep{abb+18a,ykh+21} and exotic forms of \hbindex{dark matter} \citep[e.g.][]{cyh20}. For the interested reader, the wealth of accessible science with GW limits and detections is the subject of another large review \citep{btc+19}.

For the purposes of this review, it suffices to say that if our limits continue to improve around an order of magnitude beyond their current point, it would be astrophysically surprising. This is because even the most pessimistic simulations of supermassive binary black hole evolution (where no galaxy merger results directly in a binary supermassive black hole due to inefficient inspirals) still result in signals detectable at the $h_{\rm yr}> 10^{-16}$ level \citep{bsbh18}. See also the ``Massive Black Hole Mergers'' chapter of this edition by E.~Barausse and A.~Lapi.  

In summary, there are a variety of ways in which PTAs can further gain
sensitivity and all of these ways are being explored. Essentially all
predictions conclude a detection is likely within the next few years
or at most within the next decade -- regardless of which particular
improvement is being studied. 
This is not surprising given that our
most reliable predictions on the strength of the GWB are
right up against our most constraining limit \citep{srl+15,abb+18a}, which
strongly implies a detection is bound to be imminent.

\section{Summary}

The extreme properties of the spinning neutron stars called pulsars
enable a unique experiment to detect GWs in a spectral range that is
highly complementary with other mature GW projects like LIGO and
LISA. PTAs are expected to make the first detection of nHz GWs within
the next few years and in doing so, will allow new and unprecedented
constraints on galaxy formation and evolution scenarios. The way
forward is long and hard, however, as data sets are complex and highly
heterogeneous and a variety of noise sources need to be dealt
with. Instrumental upgrades and extension and improvement of observing
schedules and source lists are underway to further enhance sensitivity
-- a process that will culminate in the ultimate PTA to be ran on the
telescope of the future: the SKA. With the added collecting area and
the larger number of pulsars that the SKA will be able to time at high
precision, GW astronomy in the nHz range can be expected to properly
take off.

\section{Cross-References}
%

Barausse \& Lapi, ``Massive Black Hole Mergers''

\begin{acknowledgement}

JPWV acknowledges support by the Deutsche Forschungsgemeinschaft (DFG)
through the Heisenberg programme (Project No.\ 433075039). SBS is
supported by NSF awards \#1458952 and \#1815664 and is a CIFAR Azrieli
Global Scholar in the Gravity and the Extreme Universe program.

\end{acknowledgement}



\bibliographystyle{spbasic}
\bibliography{}


\printindex

\end{document}